\documentclass[twocolumn,amsmath,amssymb,superscriptaddress]{revtex4-1}
\usepackage{graphicx}
\usepackage{color}
\usepackage[caption=false]{subfig}

\def\figdir{.}

\begin{document}

\title{Approximate density matrix functionals applied to hetero-atomic bond dissociation} 

\author{Robert van Meer} 
\email[Corresponding Author. Email: ]{rvanmeer@gmail.com} 
\affiliation{Department of Physics, National Taiwan University, Taipei 10617, Taiwan} 

\author{Jeng-Da Chai} 
\email[Corresponding Author. Email: ]{jdchai@phys.ntu.edu.tw} 
\affiliation{Department of Physics, National Taiwan University, Taipei 10617, Taiwan} 
\affiliation{Center for Theoretical Physics and Center for Quantum Science and Engineering, National Taiwan University, Taipei 10617, Taiwan} 

\date{\today} 

\begin{abstract} 
\noindent 
A two-orbital two-electron diatomic model resembling LiH is used to investigate the differences between the exact L\"{o}wdin-Shull and approximate Hartree-Fock-Bogoliubov and Baerends-Buijse 
density matrix functionals in the medium- to long-distance dissociation region. In case of homolytic dissociation (one electron on each atom), the approximate functionals fail to generate the correct 
energy due to a compromise between the Hartree-Fock component (which favors partial charge transfer) and the strong correlation component (which hampers charge transfer). The exact functional 
is able to generate the physically correct answer by enforcing the equi-charge distribution of the bonding and antibonding orbitals. Besides, the approximate functionals also have issues in correctly 
describing heterolytic dissociation (two electrons on one atom) due to the strong correlation component hampering charge transfer. In this work, we propose a new scheme in which the homolytic 
dissociation problem for approximate functionals is avoided by adding a Lagrange multiplier that enforces equi-charge distribution of the bonding and antibonding orbitals. The symmety based nature of 
the findings implies that they are most likely transferable to other cases in which one uses an approximate one-particle method in conjunction with a symmetrical particle-hole correction factor.
\end{abstract} 

\maketitle

\section{Introduction} 
\noindent 
Kohn-Sham density functional theory (KS-DFT) \cite{KohnSham1965} with conventional approximate exchange-correlation (XC) density functionals is incapable of describing strong (static) correlation 
processes, such as bond breaking and bond formation, so one has to go beyond this approach in the sense that additional quantities have to be used. Traditionally, these processes have been described by 
wavefunction-based methods. The most commonly used strategy for strong correlation cases is to use a variational (nearly) complete-active-space (CAS) type calculation that only has the orbitals involved in 
the strong correlation process in its active space as a starting point. The remainder of the missing dynamical correlation is then generated by various non-variational methods, such as perturbation theory and 
the use of correlation density functionals in KS-DFT \cite{Nishimoto2019,GhoshGagliardiTruhlar2018,GritsenkovanMeerPernal2018}. 

These type of approaches are routinely successfully applied to situations in which one only has a few orbitals and electrons in the active space. Unfortunately, the variational CAS part becomes the bottleneck 
for situations in which many orbitals and electrons are present in the active space, necessitating the use of configurational deadwood deselection methods that use some smart selection criteria or stochastic 
sampling \cite{LischkaBarbatti2018,BoothThomAlavi2009}. Considering the fact that variationality is lost in any case when one adds a non-variational dynamical correlation correction on top of the variational 
wavefunction, one can also consider using non-variational functional approaches for the CAS space in order to lower the computational cost. 

Within the framework of density functional theory, thermally-assisted-occupation density functional theory (TAO-DFT) \cite{Chai2012} is a very efficient method that can tackle static correlation problems. 
TAO-DFT is a density functional theory with fractional orbital occupations produced by the Fermi-Dirac distribution (controlled by a fictitious temperature that is related to the distribution of the exact natural 
orbital occupation numbers), wherein an entropy contribution term can approximately describe static correlation even when the simplest local density approximation XC density functional is adopted. More 
complicated XC density functionals, such as the generalized-gradient approximation \cite{Chai2014} and hybrid \cite{Chai2017} XC density functionals, can also be adopted in TAO-DFT. In addition, an 
approach that determines the fictitious temperature in TAO-DFT in a self-consistent manner \cite{LinHuiChungChai2017} has been recently developed to improve the overall accuracy of TAO-DFT for diverse 
applications. Recently, TAO-DFT has been adopted to study the ground-state properties of several nanosystems with pronounced radical nature \cite{TAOa1,TAOa2,TAOa3,TAOa4}. 

Among the methods that go beyond density functional theory, density matrix functional theory 
(DMFT) \cite{Gilbert1975,Goedecker1998,Muller1984,BuijseBaerends2002,Gritsenko2005a,LathiotakisSharma2009,Piris2017,PernalGiesbertz2016,TheophilouLathiotakisHelbig2016,vanMeerGritsenko2019,HollettLoos2020}, 
can also tackle static correlation problems. This method will be adopted and discussed in the present work. 

In DMFT, the ground-state electronic energy is written as a functional of the one-body reduced density matrix: 
\begin{equation} 
E_{\text{el}} = \sum_{i} 2n_{i} h_{ii} + W[\{n_{i}\}, \{\phi_{i}\}] 
\end{equation} 
here $n_{i}$ are the natural orbital occupation numbers whose value ranges from 0 to 1, $\phi_{i}$ are the natural orbitals (NOs), $h_{ii}$ are the one-electron integrals, and $W[\{n_{i}\}, \{\phi_{i}\}]$ is the 
DMFT electron-electron interaction functional. An exact L\"{o}wdin-Shull (LS) electron-electron interaction functional is known for two-electron systems \cite{LowdinShull1956}: 
\begin{equation} 
W_{\text{LS}} = \sum_{i,j} f_{i} f_{j} \sqrt{n_{i} n_{j}} L_{ij}. 
\end{equation} 
Here, $f_{i}$ are the phase factors whose value is usually set to 1 for the highest occupied NO and $-1$ for all the others, and $L_{ij}$ are the star conjugated exchange integrals that reduce to the normal 
$K_{ij}$ exchange integrals for ground-state energy evaluations. 

Unfortunately, an exact functional remains unknown for general $N$-electron systems, forcing one to use approximate functionals. Most of the recent approximate $N$-electron DMFT functionals can be 
classified as geminal functionals, meaning that they essentially divide the system into several ``separate" two-electron systems that mainly feel a mean-field Hartree-Fock (HF) like interaction of the other 
electron pairs, and use the LS functional internally. Geminal functionals can be considered as approximate seniority-zero wavefunctions \cite{BytautasScuseria2011,BoguslawskiTecmer2015}, as such they are 
incapable of describing the majority of the dynamical correlation, often only capturing 20--30\% of the total dynamical correlation \cite{RassolovXu2004,vanMeerGritsenkoBaerends2018}. They are, however, quite good 
at describing the active space of most systems. 

There are still a few caveats for the strongly correlated CAS space, when one uses the simplest of these geminal functionals, the antisymmetrized product of 
strongly-orthogonal geminals (APSG) method \cite{Rassolov2002}, such as the absence of correct local exchange between bond-broken electrons when multiple electrons end up on the same fragment in 
multi-bond dissociation cases, missing dispersive interactions between the geminals, and symmetry related issues for aromatic systems. These shortcomings have been tackled to a certain degree in various 
self-consistent and perturbative extensions of the APSG functional, such as the PNOF6-7 and ELS-D-M functionals \cite{Piris2014,Piris2017,vanMeerGritsenkoBaerends2018}. One of the problems that still remains is that geminal-based approaches have a somewhat more unfavorable scaling (N$^5$) compared to the self-consistent-field (SCF) like scaling of the first-generation functionals (N$^4$) due to the more expensive full orbital transformations that have to 
be performed for all geminal-based methods \cite{Giesbertz2016}. 

Commonly used first-generation functionals, such as the Hartree-Fock-Bogoliubov (HFB) and Baerends-Buijse (BB) 
functionals \cite{CsanyiArias2000,Muller1984,BuijseBaerends2002,CohenBaerends2002} 
\begin{align} 
W_{\text{HFB}} &= W_{\text{HF}_{\gamma}} + W_{\text{HFB}_c} \nonumber \\&= \sum_{i,j} n_{i} n_{j} (2J_{ij} - K_{ij}) \nonumber \\&- \sum_{i,j} \sqrt{n_{i} (1 - n_{i}) n_{j} (1 - n_{j})} K_{ij} \\ 
W_{\text{BB}}  &= W_{\text{HF}_{\gamma}} + W_{\text{BB}_c} \nonumber\\&= \sum_{i,j} n_{i} n_{j} (2J_{ij} - K_{ij}) \nonumber \\&+ \sum_{i,j} (n_{i} n_{j} - \sqrt{n_{i} n_{j}}) K_{ij} 
\end{align} 
are less successful than the geminal-based methods from a stability point of view. If one allows for full variational optimization the HFB functional fails to generate a large part of the dynamical correlation energy at equilibrium geometries and sometimes overcorrelates in the dissociation region, while the BB functional tends to overcorrelate at all distances. The overcorrelation of the BB functional was remedied by adding a successive series of repulsive corrections, resulting in the BBC3/AC3 functionals, which are capable of describing both dynamical and strong correlation for systems in which a single bond is broken. Unfortunately, the additional corrections severely affect the computational scaling \cite{Giesbertz2016}. 

Another option to avoid overcorrelation is to severely restrict the number of orbitals that are included in the correlated description. Such approaches tend to give an accurate energetic description of strong correlation in the dissociation limit for symmetrical systems, but miss a large part (if not all) of the dynamical correlation. These minimal expansions are often used in model systems, and are also seen as a potential practical approach of adding strong correlation to approximate DFT functionals. 

Unfortunately, even for simple two electron two orbital systems approximate DMFT functionals can still have some issues. For instance it has been shown that a hubbard dimer model using the BB functional allows for the mixing of singlet and triplet solutions, effectively allowing for a wide range of magnetizations \cite{KamilSchadePruschkeBlochl2016}. The largest amount of problems, including artificial energy lowering and charge transfer, have been reported for the dissociation limit description for heteroatomic systems \cite{ScuseriaTsuchimochi2009,CohenMoriSanchez2016,vanMeerGritsenko2019,HellgrenGould2019}. In ref \cite{ScuseriaTsuchimochi2009} it is stated that in case of dissociation fragment orbital localization for the HFB functional the difference in chemical potential is responsible for the charge transfer. A more recent addition \cite{HellgrenGould2019}, in which a local potential form of the BB is used \cite{LathiotakisHelbig2014PRA,LathiotakisHelbig2014JCP}, it is argued that the local BB functional is suffering from charge transfer due to deficiencies in the effective potential compared to the exact one. So the reasons behind these failures are not fully understood yet in a comprehensive way. It is interesting to do a more thorough analysis of the various energy components to see what causes this behavior, and whether one can take advantage of both the computational scaling and better energetics at the same time by resolving the problem. 

To this end, we have used a simple hetero-diatomic two-orbital two-electron model. Such a model has been used by many different research groups in the past to understand the behavior of (approximate) TDDFT \cite{FuksMaitra2014,CarrascalFerrerMaitraBurke2018}, DFT \cite{TempelMartinezMaitra2009,HelbigTokatlyRubio2009,CarrascalFerrerSmithBurke2015} and also RDMFT \cite{ScuseriaTsuchimochi2009,CohenMoriSanchez2016,MitxelenaPirisRodriguez2017,HellgrenGould2019} compared to the exact two electron wavefunction system. The main unique point of our analysis is the fact that we fully elucidate the reason behind the combined charge transfer, orbital rotation and occupation number errors for the BB and HFB functionals and attribute this to the structure of the Hartree Fock and correlation correction components. 

In Section II, the full details of the model are described. Section III shows and compares the results of the various functionals that have been tested. A possible solution to the heteroatomic dissociation problem is suggested at the end of Section III. The conclusions are drawn in Section IV.

\section{Model Details} 

We are going to use a simple two-orbital two-electron diatomic model to investigate the differences between the exact two-electron DMFT functional and commonly used approximate functionals. A two-orbital 
model will not be capable of describing a large part of the dynamical correlation, so we will only look at the medium- to long-range dissociation limit for which the model description will be close to the 
non-model case. At this distance, the overlap of atomic orbitals $A$ and $B$ can be assumed to be negligible. The left ($l$) and right ($r$) leaning (natural/molecular) orbitals for such a case are given by 
\begin{align*} 
\phi_{l} = \frac{1}{\sqrt{1 + \lambda}}\left(A + \sqrt{\lambda} B\right) \\ 
\phi_{r} = \frac{1}{\sqrt{1 + \lambda}}\left(\sqrt{\lambda} A - B\right) 
\end{align*} 
here $0 \leq \lambda \leq 1$ is a mixing parameter. These orbitals reduce to their respective atomic orbitals at $\lambda = 0$, while one obtains homopolar orbitals at $\lambda = 1$ \cite{Surjan1999}. The 
intermediate values of $\lambda$ allow one to interpolate between these two extreme cases, and generate all possible charge-distribution ratios. Using these orbitals and the neglect of overlap, we get the 
following expressions for the required integrals 
\begin{align*}
h_{ll} &= \frac{h_{AA} + \lambda h_{BB}}{1 + \lambda} - \frac{1}{R} \\ 
h_{rr} &= \frac{\lambda h_{AA} + h_{BB}}{1 + \lambda} - \frac{1}{R} \\ 
J_{ll} &= \frac{1}{(1 + \lambda)^2}\left(J_{AA} + \lambda^{2} J_{BB} + \frac{2 \lambda}{R}\right) \\ 
J_{rr} &= \frac{1}{(1 + \lambda)^2}\left(\lambda^{2} J_{AA} + J_{BB} + \frac{2 \lambda}{R}\right) \\ 
J_{lr} &= \frac{\lambda}{(1 + \lambda)^2}\left(J_{AA} + J_{BB}\right) + \frac{1 + \lambda^{2}}{(1 + \lambda)^2}\frac{1}{R} \\ 
K_{lr} &= \frac{\lambda}{(1 + \lambda)^2}\left(J_{AA} + J_{BB} - \frac{2}{R}\right). 
\end{align*} 
Here, $h_{AA}$ and $h_{BB}$ are the atomic one-electron integrals, and $J_{AA}$ and $J_{BB}$ are the atomic two-electron Coulomb repulsion integrals, and $R$ is the distance between the atoms. 
In addition to the orbitals, DMFT also uses the natural orbital occupation numbers directly in order to generate prefactors for the integrals. In principle, we have two occupation numbers $n_{l}$ and $n_{r}$. 
However, the sum of these occupations should be 1 (i.e., for two electrons), so we can express both of them using a single variable 
\begin{align*} 
n_{l} &= 1 - x \\ 
n_{r} &= x. 
\end{align*} 
Using these occupation number expressions, we obtain the following functional energy expressions for our model system 
\begin{align*} 
E_{\text{HF}_\gamma} &= 2((1 - x) h_{ll} + x h_{rr}) + (1 - x)^2 J_{ll} + x^2 J_{rr} \\&+ x(1 - x)(4J_{lr} - 2K_{lr}) + \frac{1}{R} \\ 
E_{\text{HFB}} &= 2((1 - x) h_{ll} + x h_{rr}) + (1 - x)^2 J_{ll} + x^2 J_{rr} \\&+ x(1 - x)(4J_{lr} - 2K_{lr}) \\&- x(1 - x)(J_{ll} + J_{rr} + 2K_{lr}) + \frac{1}{R} \\ 
E_{\text{BB}} &= 2((1 - x) h_{ll} + x h_{rr}) + (2 (1-x)^2 - (1-x)) J_{ll} \\&+ (2 x^2 - x) J_{rr} + 4x(1 - x) J_{lr} \\&- 2 \sqrt{x(1 - x)} K_{lr} + \frac{1}{R} \\ 
E_{\text{LS}} &= 2((1 - x) h_{ll} + x h_{rr}) +(1 - x) J_{ll} + x J_{rr} \\&- 2 \sqrt{x(1 - x)} K_{lr} + \frac{1}{R} 
\end{align*} 
here the final $\frac{1}{R}$ at the end of each energy expression represents the effective nuclear repulsion. It should be noted that the majority of the geminal type functionals (e.g. PNOF5, PNOF7 and ELS-D-M) reduce to the exact LS functional for our model system, so there is no need to investigate them separately. As can be seen from the energy expressions, only singlet states are studied in our model.

In principle, one can try to understand the differences between the functionals by transforming this model into an asymmetrical two-site Hubbard system \cite{CohenMoriSanchez2016}. However, it is more 
fruitful to simply generate grid-based figures (2D, $x$ and $\lambda$) for a practical system, since this allows us to gauge the importance of different components, and focus our attention to the important 
ones for a real system. The system of choice is the LiH molecule. This molecule is one of simplest hetero-diatomic molecules whose valence electronic structure is essentially a two-electron system, allowing 
us to still use the LS functional while avoiding the symmetry which is present in the often used H$_{2}$ prototype molecule. We have used the \textsf{GAMESS-US} program \cite{GAMESS} to generate the 
numerical values for the atomic orbital quantities. The values were obtained from restricted open-shell Hartree-Fock (ROHF) calculations that use the cc-pVTZ basis (spherical). The $2s$ one-electron and 
self-repulsion energies for the Li atom (i.e., atom $B$) are $-$0.19631 (includes core 2J-K interaction) and 0.2341 hartree. The $1s$ quantities for the H atom (i.e., atom $A$) are $-$0.4998 and 0.6251 hartree, 
which is close to the exact values of $-$0.5 and 0.625 hartree, respectively. The frozen-core energy of the Li atom ($-$7.23637 hartree) has been omitted from the calculations. This does not alter the shape of 
plots, since its contribution is uniform across the grid. 

In addition to the occupation ($x$) and orbital ($\lambda$) mixing parameters, there is still a third (somewhat hidden) variable, namely the distance $R$ between the atoms. The long-distance component is 
generally avoided in most studies that evaluate the dissociation behavior of functionals. This is, however, not a very wise choice. In case of charge transfer, erroneous or physically motivated, the long-distance 
component will stabilize the unequal charges, and have a large impact on the location of the minimum on the grid. In addition to this, unbeknownst to many, even with equal atoms one can still end up having 
a non-zero long-distance interaction for certain occupation number scenarios, if one uses the HF functional. 

\begin{figure}[!p] 
\includegraphics[width= 0.5\textwidth, trim= 20 20 20 20]{\figdir/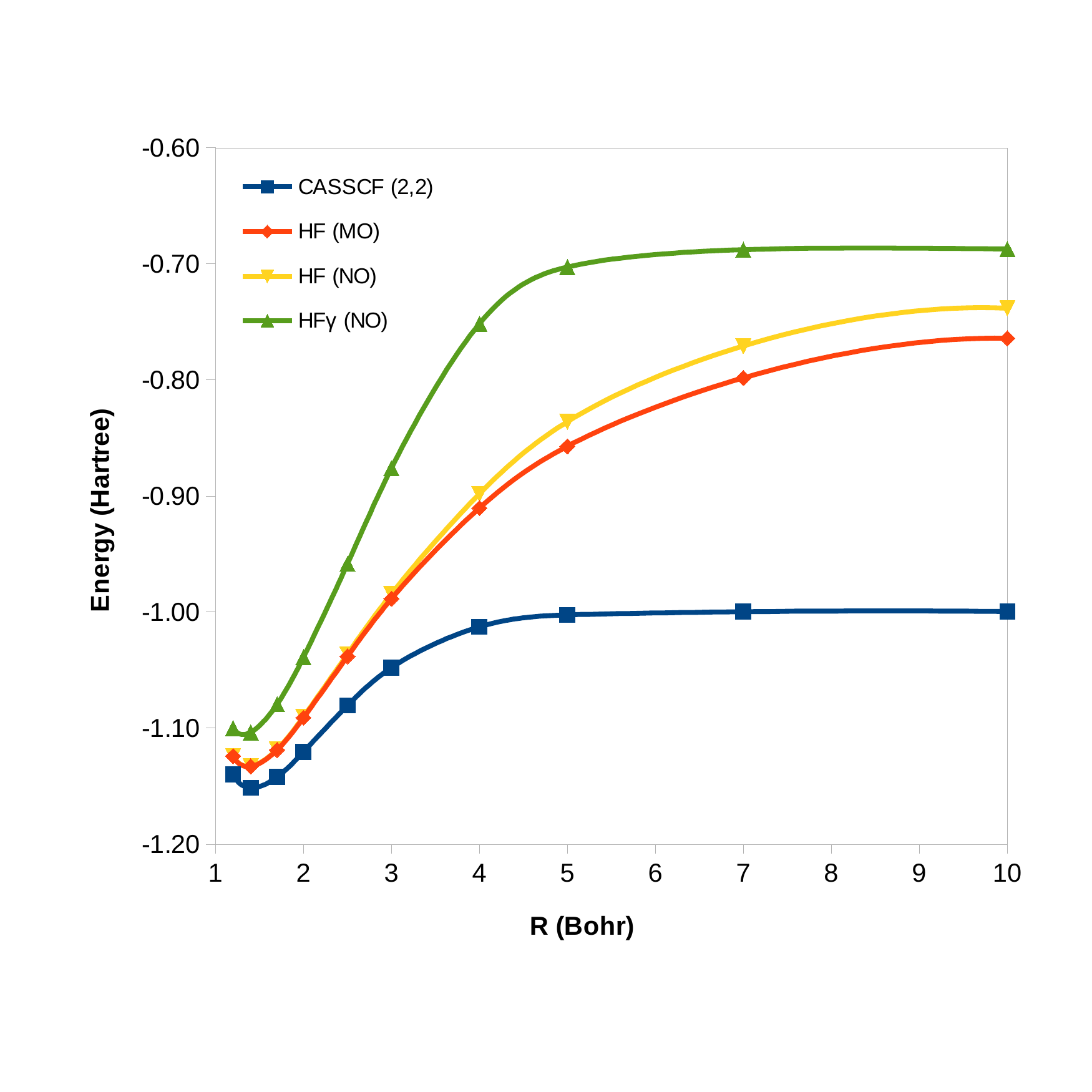} 
\caption{\label{fig:H2HFplot} 
Energy curve of H$_{2}$ for CASSCF(2,2) (i.e., the complete-active-space self-consistent-field method that includes 
two active electrons distributed among two active orbitals) and several Hartree-Fock methods. 
HF$_\gamma$ (NO): HF functional using the occupations and NOs obtained from CASSCF(2,2) calculations. 
HF (NO): HF functional using the highest occupied NO from CASSCF(2,2) calculations and an aufbau occupation. 
HF (MO): Conventional aufbau Hartree-Fock using a doubly occupied SCF generated MO. 
All calculations use the Cartesian cc-pVTZ basis. 
The energy difference between the HF (NO) and HF (MO) approaches is related to the latter being able to relax the orbital to 
accommodate the additional local (erroneous) repulsion \cite{BuijseBaerendsSnijders1989,BuijseBaerends2002}.} 
\end{figure}

A simple example is the H$_{2}$ molecule ($\lambda = 1$, atom $A$ = atom $B$). For the standard 
aufbau solution, the Fermi hole generated by the exchange integral is spread across both atoms, so at each atom, an electron feels the field of half an electron at the same atom, and another half of an electron 
at the other atom. The distant atom still has a nuclear charge of 1 ($\frac{1}{R}$ nuclear repulsion, $-\frac{2}{R}$ electron-nucleus attraction), so the total long-distance interaction is an attractive $-\frac{1}{2R}$. 
This attractive interaction is responsible for the slow Hartree-Fock convergence towards the dissociation asymptote (Figure \ref{fig:H2HFplot}). When one transfers some fractional charge from the bonding to 
the antibonding orbital, the local charge interaction of the aufbau solution is retained. However, the long-distance component is altered due to the presence of off-diagonal exchange integrals $K_{lr}$ whose 
long-distance component flips its sign with respect to the Coulomb entries due to the phase (+/$-$ sign) interaction of the different molecular/natural orbitals. In case of equal occupation of the bonding and 
antibonding orbitals (i.e., $x$ = 0.5), the long-distance component of the Hartree-Fock two-electron interaction reduces completely to the exact one \cite{CohenBaerends2002} 
\begin{align} 
W_{\text{HF}_{\gamma}} = \frac{1}{4} J_{ll} + \frac{1}{4} J_{rr} + J_{lr} - \frac{1}{2} K_{lr} = \frac{1}{4} (J_{AA} + J_{BB}) + \frac{1}{R}. 
\end{align} 
So the total long-range interaction vanishes for this situation, as does the long-distance ionic tail in the energy plot (Figure \ref{fig:H2HFplot}). This means that non-aufbau HF solutions have a higher energy 
than the aufbau solutions (with the same value of $\lambda$) for all non-zero $\frac{1}{R}$ values. The fact that the aufbau HF solutions have a lower energy than the non-aufbau HF was already proven theoretically by Lieb \cite{Lieb1981}. It is instructive to see how this takes shape in the dissociation region.

In order to take into account and study the effects of the long-distance component, we have performed all calculations for 4 different choices of the distance. The first distance is infinity, which removes all 
distance-related effects, and is essentially the true dissociation limit. The next choices are 10 and 5 bohr, yielding $\frac{1}{R}$ values of 0.1 and 0.2 hartree, respectively. These distances represent the 
type of distances between which one generally halts the potential energy curve calculation, since most curves reach their dissociation limit asymptote in this region. All 10 bohr plots have been relegated to the 
supporting information section in order to save space, the 5 bohr plots paint a more exaggerated picture without altering the physics. The last value is somewhat special in the sense that the system starts to 
exhibit heterolytic dissociation (ion formation), instead of homolytic dissociation (one electron on each atom). The point at which heterolytic dissociation becomes more favorable (assuming 
$2h_{AA} + J_{AA} < 2h_{BB} + J_{B}$, i.e. it is more favorable to put two electrons on atom A than on atom B) is given by 
\begin{align} 
2h_{AA} + J_{AA} - \frac{1}{R} \leq h_{AA} + h_{BB}. 
\end{align} 
Here the $-\frac{1}{R}$ term represents the ionic long distance ``stabilization" energy. For our LiH model system, this occurs when $\frac{1}{R} \geq 0.322$ hartree. We have performed calculations at $\frac{1}{R} = 0.32$ hartree (just before the jump), in order to see how the energy surface accommodates the switch from one minimum to another. It should be noted that in all cases, we are assuming that there is no overlap between the atoms. This is not an issue for infinite distance, and is most likely not a problem for 10 and 5 bohr. However, it is very unphysical for the final choice. So one should view this case as interesting model scenario. In reality, a similar scenario could occur at more 
physical distances in case $2h_{AA} + J_{AA}$ is much more favorable with respect to $h_{AA} + h_{BB}$ than it is for our current LiH system. 

In this work, we have not allowed for any relaxation of the atomic orbitals, and we are only using two orbitals. This is essentially correct for the exact homolytic dissociation case. However, this is not true for 
heterolytic scenario. In order to avoid local repulsion, the actual highly occupied natural orbital is going to be more diffuse than the one that we are using. In addition to this, more orbitals are required to fully 
describe the dynamical correlation. So in reality, the point at which heterolytic dissociation becomes viable should be slightly lower than the value that we are using in our model. These model shortcomings 
should have very little to almost no impact on the ideas obtained from our findings.

\section{Results} 

In this section, we are going to evaluate the LiH model 2D energy grid for several DMFT functionals and remedy the problematic homolytic dissociation for the approximate functionals. The global minima of 
each plot is shown in Table \ref{tab:ModelMinima}. We have assigned $A$ as the Hydrogen atom and $B$ as the Lithium atom. As a reminder, both orbitals contain an equal amount of charge on each atom at 
$\lambda = 1$, while each orbital has all of its charge on a single atom at $\lambda = 0$. When $x < 0.5$ and $\lambda < 1$, there is more charge on the Hydrogen atom; when $x > 0.5$ and $\lambda < 1$, 
there is more charge on the Lithium atom.

\subsection{Hartree-Fock functional} 

\begin{figure}[!p] 
\subfloat[]{\includegraphics[width= 0.5\textwidth, trim= 20 20 20 20]{\figdir/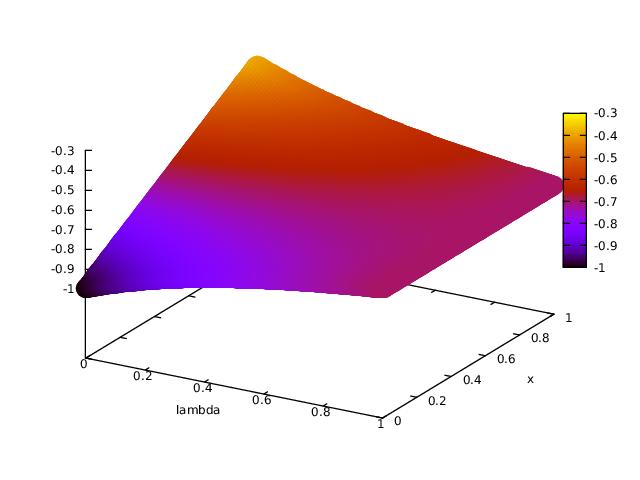}} \\ 
\subfloat[]{\includegraphics[width= 0.5\textwidth, trim= 20 20 20 20]{\figdir/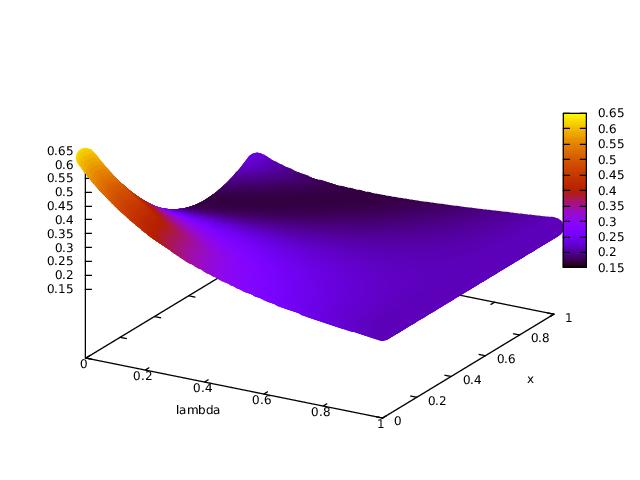}} 
\caption{\label{fig:eHF1el2elgrid} 
Hartree-Fock one-electron and two-electron energy contribution grid for $\frac{1}{R}$ = 0 (hartree).} 
\end{figure}

We will begin our analysis by looking at the infinite-distance HF results. In order to gain some more understanding, we are first going to observe the plots of the one-electron part (Figure \ref{fig:eHF1el2elgrid}a) 
and the two-electron part (Figure \ref{fig:eHF1el2elgrid}b), separately. The one-electron plot shows that it is most favorable to put both electrons on the H atom ($x = 0$, $\lambda = 0$), and least favorable to 
not put any charge on the Hydrogen atom ($x = 1$, $\lambda = 0$), which is easily understood if one looks at the one-electron orbital energies. At $\lambda = 1$, the plot is symmetrical (flat) along the 
$\lambda = 1$ line. This can be explained by the fact that there is essentially no difference between the one-electron energies of the molecular orbitals for $\lambda = 1$, so any combination of occupations is 
going to yield the same sum of atomic one-electron contributions. In addition to this symmetry, the entire line $x = 0.5$ has the same energy regardless of $\lambda$, which is again caused by having the same 
atomic one-electron contributions due to symmetry considerations. It should be mentioned that these effects are also valid for all other functionals, since the one-electron part is universal. Now we will shift 
towards looking at the two-electron HF contributions. The plot (Figure \ref{fig:eHF1el2elgrid}b) clearly shows that it is more beneficial to put more charge on the Li atom ($x > 0.5$, $\lambda < 1$), but not 
everything. The preference for putting more charge on the Li atom can easily be explained by the smaller self-repulsion integral. The reason for not putting all charge on the Li atom is that the repulsion does not 
scale linearly. In addition to these preferences, the HF two-electron part has the same T-shape symmetry (along $x = 0.5$ and $\lambda = 1$) as the one-electron part. So the entire HF energy has a T-shape 
symmetry, if one does not take into account any long-range interactions. 

\begin{figure}[!p] 
\subfloat[]{\includegraphics[width= 0.5\textwidth, trim= 20 20 20 20]{\figdir/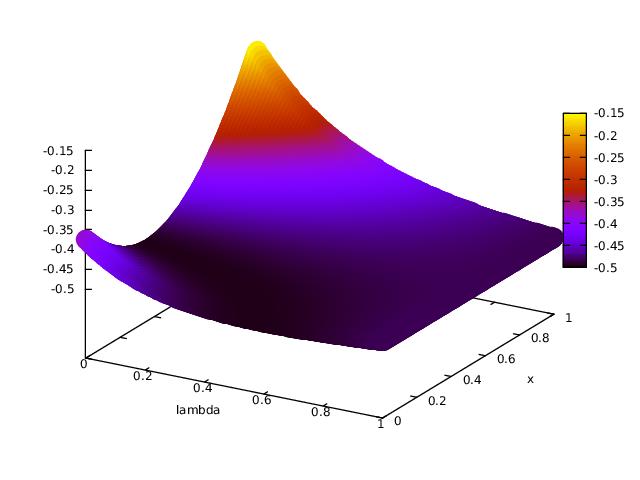}} \\ 
\subfloat[]{\includegraphics[width= 0.5\textwidth, trim= 20 20 20 20]{\figdir/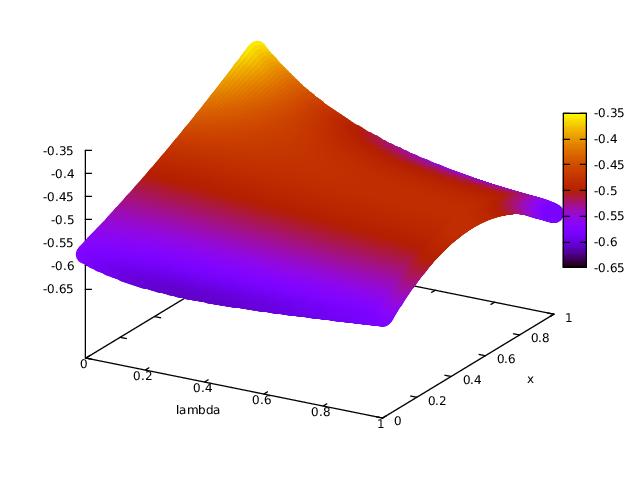}} \\ 
\subfloat[]{\includegraphics[width= 0.5\textwidth, trim= 20 20 20 20]{\figdir/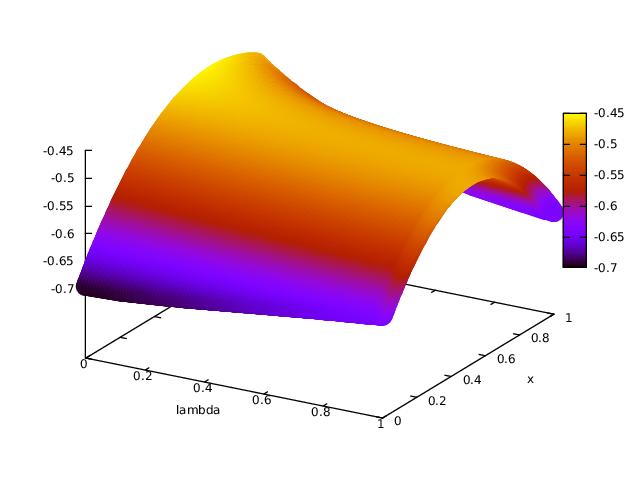}} 
\caption{\label{fig:eHFgrid} 
Hartree-Fock energy grid for $\frac{1}{R}$ (in hartree) = 0 (a), 0.2 (b) and 0.32 (c).} 
\end{figure}

We are now going to look at the full plot of the combined result. The plot (Figure \ref{fig:eHFgrid}a) shows that the minima is located in the $x < 0.5$ and $\lambda < 0.6$ region, which essentially means that 
the one-electron terms are more important than the two-electron terms, and that the system prefers putting more charge on the Hydrogen atom (partial heterolytic/charge transfer bond break). It is interesting to 
note that the actual minimum is located at $x = 0.26, \lambda = 0.286$, which indicates that the HF functional has a non-aufbau minimum. It should be mentioned that there is an entire energy groove with almost 
the same energy around the minimum, so the actual location of the minimum might still be an aufbau solution ($x = 0.0, \lambda = 0.61$), if all potential numerical issues are taken into account. Nonetheless, it is 
quite interesting to see such an energy groove. Its existence can be explained by the fact that one essentially maintains the same local one-electron terms and repulsions, if one carefully alters the occupation 
and molecular orbital composition at the same time. 

Now that we have seen the behavior of the HF functional without long-distance effects it is time to include said effects. We will start by looking at the 5 bohr plot (Figure \ref{fig:eHFgrid}b). It is quite clear that the 
previously mentioned energy groove has disappeared completely. And also, the T-shape symmetry has been altered. The $x = 0.5$ symmetry remains, but the $\lambda = 1$ symmetry is changed from a flat line 
to a symmetrical parabola-like shape. Both of these findings can explained by the additional long-distance stability that HF gains when using aufbau occupations. As explained above in case of an aufbau 
solution, the Fermi hole is distributed among both atoms, resulting in a net attractive long-distance interaction. If one mixes the occupations, the off-diagonal exchange ($K_{lr}$) terms will start cancelling this 
attractive term, reaching a full cancellation at $x = 0.5$ for any $\lambda$. This explains the parabolic shape at $\lambda = 1$ and the removal of the energy groove in the $x < 0.5$ and $\lambda < 0.5$ region 
(it is energetically more favorable to select the aufbau solution with the same charge distribution in order to gain an unphysical total attractive long-distance interaction). It should be mentioned that the parabolic 
shape is not maintained for all $\lambda$. At $\lambda = 0$ the long distance behavior is completely governed by the $J_{lr}$ integral, and no unphysical long distance stabilization can take place. Instead, the near linear switching (one-electron terms) between the two ionic configurations dominates, resulting in linear behavior when moving away from $x = 0.5$. This does not directly lead to a HF minimum along $\lambda = 0$, but it does play an important role for the minima of functionals that are (somewhat) based on HF, namely the HFB and BB functionals, since one can obtain the highest amount of energy gain when moving away from the $x = 0.5$ line at $\lambda = 0$. The actual HF minimum is a proper aufbau minimum located at $x = 0.0, \lambda = 0.36$. If one compares this minimum with the aufbau minimum of the non-long-distance plot, one can see that additional attractive long-distance interaction promotes more charge transfer to the heterolytically favored atom (e.g., H atom in our case). 

Stronger long-distance attractive terms lead to more charge transfer. In case of the pre-heterolytic bond dissociation point (Figure \ref{fig:eHFgrid}c), one can see that the minimum is shifted towards $x = 0.0, \lambda = 0.01$. Beyond the heterolytic dissociation point ($\frac{1}{R} = 0.322$ hartree), the minimum will facilitate full transfer of both electrons to the Hydrogen atom. It is interesting to note that the linear behavior (with respect to the $x$) that was observed for $\lambda = 0$ at 5 bohr has now been replaced by quadratic-like behavior around $x=0.5$. This shape can be explained by the fact that the $\frac{1}{R}$ term present in the $J_{lr}$ integral whose prefactor $4x(1-x)$ maximizes at $x = 0.5$ starts to become more dominant for larger $\frac{1}{R}$ values.

\subsection{Hartree-Fock-Bogoliubov functional} 

The HF results were quite instructive; however in all non-heterolytic dissociation cases, HF is not capable of giving an adequate description of the system. Its minima represent partial heterolytically dissociating 
systems with unphysical long-distance attractive terms. Even when the one-electron component is correctly described and the long-distance terms are corrected ($x = 0.5$), there is still an unphysical local 
repulsion. 

\begin{figure}[!p] 
\includegraphics[width= 0.5\textwidth, trim= 20 20 20 20]{\figdir/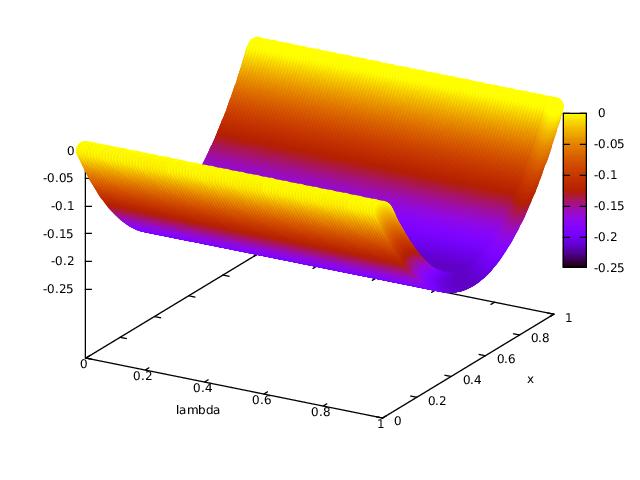} 
\caption{\label{fig:eHFBmHF} 
Hartree-Fock-Bogoliubov strong correlation component energy grid. The correction is independent of $R$ and $\lambda$.} 
\end{figure}

\begin{figure}[!p] 
\subfloat[]{\includegraphics[width= 0.5\textwidth, trim= 20 20 20 20]{\figdir/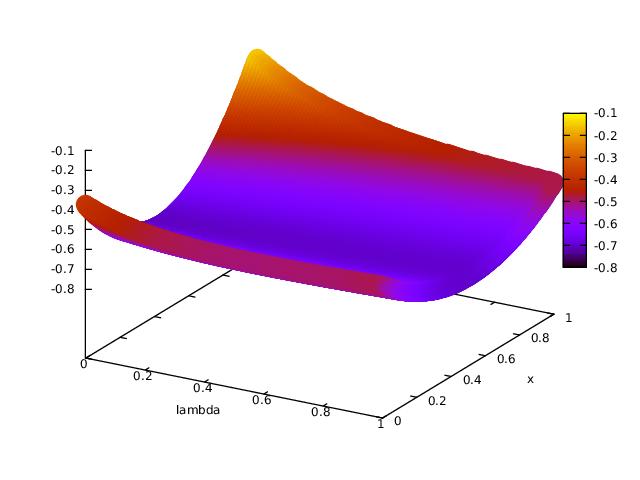}} \\ 
\subfloat[]{\includegraphics[width= 0.5\textwidth, trim= 20 20 20 20]{\figdir/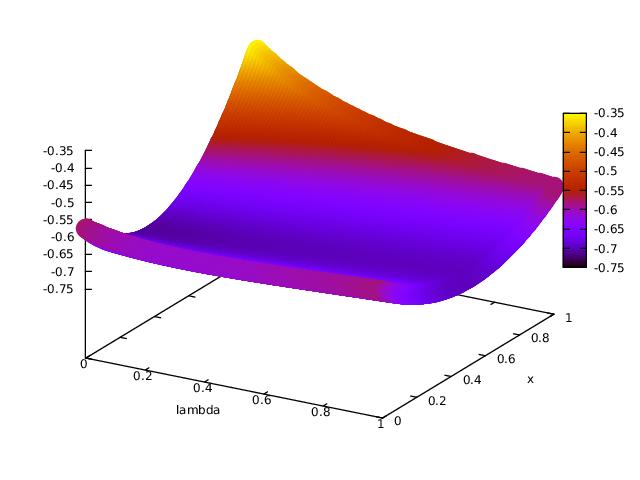}} \\ 
\subfloat[]{\includegraphics[width= 0.5\textwidth, trim= 20 20 20 20]{\figdir/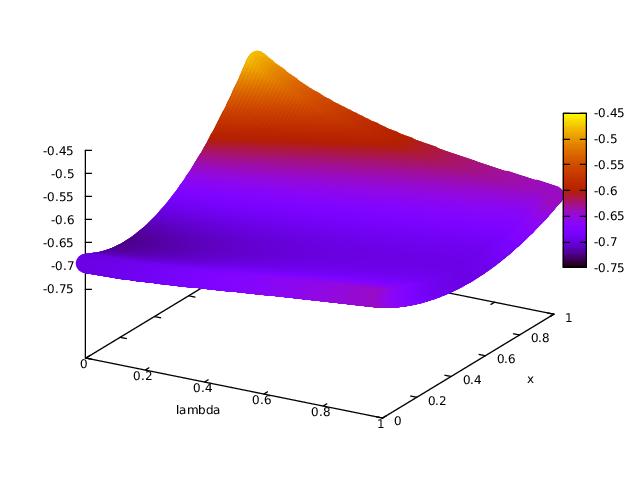}} 
\caption{\label{fig:eHFBgrid} 
Hartree-Fock-Bogoliubov energy grid for $\frac{1}{R}$ (in hartree) = 0 (a), 0.2 (b) and 0.32 (c).} 
\end{figure}

The HFB functional is ought to be capable of fixing these issues, under the right circumstances. For our model system, it adds the $J_{ll} + J_{rr} + 2 K_{lr}$ integrals (with a symmetrical $-x (1 - x)$ 
occupation-number-dependent prefactor) to the HF energy. It is interesting to write down the more explicit version of this sum 
\begin{align} 
J_{ll} + J_{rr} + 2 K_{lr} = \frac{\lambda^{2} + 2 \lambda + 1}{(1 + \lambda)^{2}}\left(J_{AA} + J_{BB}\right) \nonumber\\= J_{AA} + J_{BB}. 
\end{align} 
As one can see, it generates a fixed value, which is independent of $\lambda$, and also does not contain any long-distance effects. The previously mentioned prefactor is always symmetrical along the $x = 0.5$ 
axis, so the HFB energy contribution terms will result in a $\lambda$-less parabolic energy shift (Figure \ref{fig:eHFBmHF}), whose extremum provides the most energy reduction. We can now use the shape of 
the additional HFB contributions and the stand-alone HF curves to explain the HFB plot (Figure \ref{fig:eHFBgrid}a). At $x = 0.5$, the total HFB energy correctly describes a homolytically dissociated system 
(independent of $\lambda$). However, this line is only the minima of the additional HFB contribution, the HF curves which are a part of the total HFB energy do not have minima in this region, but instead favor 
the $x < 0.5, \lambda < 0.6$ region. The additional HFB terms start to deteriorate quite rapidly, when one moves away from $x = 0.5$. So, the total HFB minimum is a comprise of these conditions, and lies on the 
$x < 0.5, \lambda = 0$ line segment. At this point, one reaps a large benefit from moving towards the HF energy groove part, and just loses a minimal amount of the additional HFB energy. One should keep in 
mind that this minimum is not a properly homolytically dissociated system, but a compromise between HF and an energy correction which is only fully valid for $x = 0.5$. So, in essence, HFB also suffers from 
artificial charge transfer for heteroatomic systems. 

Just like for HF adding long-distance interaction (Figure \ref{fig:eHFBgrid}b) will move the minimum more towards a point where there is more charge on the 
Hydrogen atom ($x$ at the minimum changes from 0.438 to 0.382, when setting the distance at 5 bohr). Near the heterolytic switch-over point (Figure \ref{fig:eHFBgrid}c), the minimum is still quite far away from 
the fully ionic solution at $x = 0, \lambda = 0$. Even at $\frac{1}{R} = 0.35$ hartree, which is quite far beyond the switching value of 0.322 hartree, the HFB minimum still describes a system with partial 
charge transfer. Only at even larger values of $\frac{1}{R}$ does the HFB minimum describe a fully ionic distribution. This behavior is essentially caused by the same thing that is causing the erroneous homolytic 
dissociation, namely HFB is compromise of two components, a HF component and a symmetrical correction factor. In case of homolytic dissociation, the HF part forces a solution away from the physically correct 
$x = 0.5$ region. The reverse occurs for the heterolytic case, the HFB correction factor forces a minimum away from the physically correct HF ionic configuration.

\subsection{Baerends-Buijse functional} 
One can view the BB functional as a correction to the HF part that can be rewritten 
\begin{align} 
E_{\text{BB}} &= E_{\text{HF}_\gamma} - x (1 - x) (J_{ll} + J_{rr}) \nonumber\\& + 2 (x (1 - x) - \sqrt{x (1 - x)}) K_{lr} 
\nonumber\\ &= E_{\text{HF}_\gamma} - x (1 - x) (J_{ll} + J_{rr} + 2 K_{lr}) \nonumber\\& - 2 (\sqrt{x (1 - x)} - 2 x (1 - x)) K_{lr} 
\nonumber\\ &= E_{\text{HFB}} - 2 (\sqrt{x (1 - x)} - 2 x (1 - x)) K_{lr}. 
\end{align} 
So, for our model system, one can also view the BB functional as a HFB functional with an additional correction term. This correction term only has a non-zero contribution away from the $x = 0.5$ line (and the 
aufbau regions), and the $\lambda = 0$ axis (E-shape 0). The negative semi-definite integral prefactor $-2(\sqrt{x (1 - x)} - 2 x (1 - x))$ remains relatively small near $x = 0.5$, so BB is expected to yield homolytic 
dissociation results that are quite close to the HFB results. The integral $K_{lr}$ of the correction term is $\lambda$-dependent, the sign of the integral depends on the relative size of the atomic repulsion 
integrals versus the long-distance component $J_{AA} + J_{BB} - \frac{2}{R}$. For small $\frac{1}{R}$ values, the integral is positive, so the total correction favors a larger $\lambda$ value. The reverse is true 
when $\frac{1}{R}$ is large. 

The results (Table \ref{tab:ModelMinima}) show that all BB minima are located along the $x < 0.5, \lambda = 0$ line segment, indicating that the HFB component dominates the 
solution. At $\lambda = 0$, BB coincides with HFB due to the vanishing BB correction factor, so there is no need for further discussion of the individual results. 

As a final note, one should keep in mind that the (near) equivalence of HFB and BB is only true for the dissociation limit. At equilibrium distance, the BB functional tends to describe a very large amount of the 
dynamical correlation, while the HFB functional essentially reduces to an HF aufbau solution entirely.

\subsection{L\"{o}wdin-Shull functional} 

\begin{figure}[!p] 
\subfloat[]{\includegraphics[width= 0.5\textwidth, trim= 20 20 20 20]{\figdir/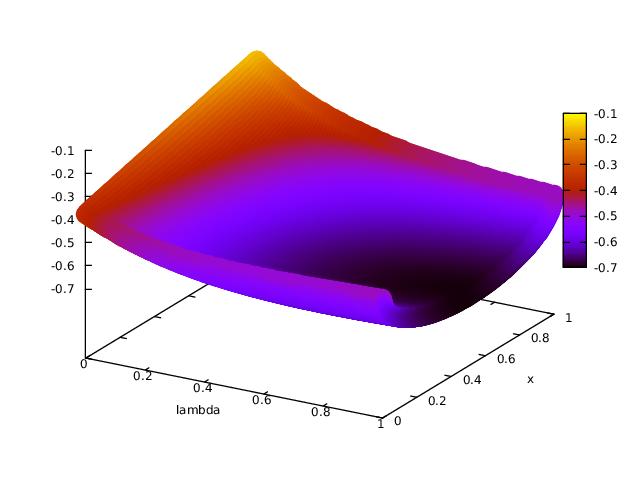}} \\ 
\subfloat[]{\includegraphics[width= 0.5\textwidth, trim= 20 20 20 20]{\figdir/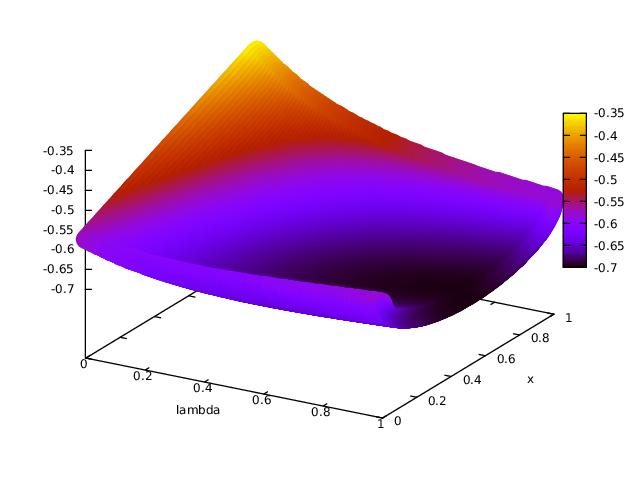}} \\ 
\subfloat[]{\includegraphics[width= 0.5\textwidth, trim= 20 20 20 20]{\figdir/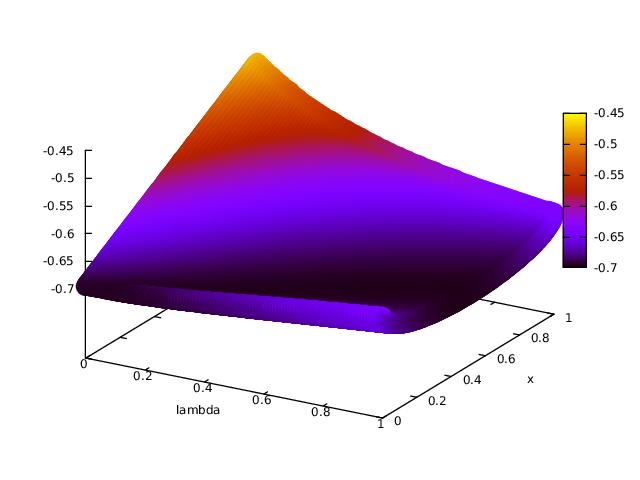}} 
\caption{\label{fig:eLSgrid} 
L\"{o}wdin-Shull energy grid for $\frac{1}{R}$ (in hartree) = 0 (a), 0.2 (b) and 0.32 (c).} 
\end{figure}

We have now finally arrived at the exact two-electron description. The LS functional is very much unlike the other functionals in the sense that it will favor a $\lambda = 1$ solution for homolytically dissociating 
systems, while still maintaining the option to correctly describe heterolytic dissociation if the need arises. It is more difficult to provide an analysis like the one that was provided for the other non-HF functionals, 
since the LS correction form is not fully symmetrical, which also explains why it generates different locations of the minima. Along the $x = 0$ and $x = 1$ edges, it behaves exactly like the HF functional, which 
it has in common with the other functionals. Just like the HFB/BB functionals, it generates a local repulsive energy removal correction that is maximal at $x = 0.5$. Unlike the other functionals, this correction is 
strongly $\lambda$-dependent for most values of $x$. At $x = 0.5$, the full two-electron component of LS is proportional (prefactor of 0.5) to 
\begin{align} 
J_{ll} + J_{rr} - 2 K_{lr} = \frac{(\lambda - 1)^{2}}{(1 + \lambda)^{2}}\left (J_{AA} + J_{BB}\right) + \frac{8 \lambda}{(1 + \lambda)^{2}} \frac{1}{R}. 
\end{align} 
As long as the self-repulsion integrals are more important than the long-distance behavior, one will need to go to $\lambda = 1$ to avoid all the self-repulsion integrals, and to obtain the proper description for a 
homolytic dissociation. At $\lambda = 1$, one naturally gets (due to symmetry) that the optimal value of $x$ is 0.5. The plot (Figure \ref{fig:eLSgrid}a) for the long-distance-less variant clearly shows the 
minimum at $x = 0.5, \lambda = 1$. The same picture is maintained for the 5 bohr results (Figure \ref{fig:eLSgrid}b). The near heterolytic dissociation plot (Figure \ref{fig:eLSgrid}c) shows that another minimum 
is forming at $x = 0, \lambda = 0$, and that there is groove between this minimum and homolytic dissociation one, showing that LS works for all two-electron systems (as it should). 

So, to summarize, the other non-HF functionals use a HF part which favors a minimum in the $x < 0.5, \lambda < 0.6$ region, and a symmetrical correction part that (mainly) favors the $x = 0.5$ line. A compromise of these two 
components leads to minima on the $x < 0.5, \lambda = 0$ line segment (for some cases, one might still get a minimum with $\lambda$ slightly larger than 0, especially for BB), whose energy is lower than the 
physically correct homolytic dissociation value. The symmetrical correction factor also inhibits the proper switch towards heterolytic dissociation. The LS functional on the other forces a homolytically dissociating 
system to go to $\lambda = 1$, and gets the correct occupation numbers due to the local symmetry. Furthermore, it allows for the creation of a corridor in the $x < 0.5, \lambda < 1$ region, if a heterolytic 
dissociation becomes viable due to the strong (linear) asymmetry of the two-electron integral prefactors when moving away from $x = 0.5$. 

The main reason for the different $\lambda$ behavior between the HFB and BB functionals on one hand, and the LS functional on the other hand is related to the role of the Coulomb integrals $J_{ll}$, $J_{rr}$, 
and $J_{lr}$ in these functionals on the $x = 0.5$ line. In case of the HFB and BB functionals, the prefactors of the diagonal terms go to zero, while both prefactors reduce to 0.5 for the LS functional. This means 
that the diagonal terms always generate electron-electron repulsion for the LS functional, while they are absent for the HFB and BB functionals. The off-diagonal Coulomb integral has a prefactor of 1 for 
the HFB and BB functionals, while it is absent for the LS functional. The off-diagonal exchange interaction is essentially the only thing that they have in common. So, the non-zero contributions will have the 
following forms 
\begin{align} 
W_{\text{LS}}^{x = 0.5} = & \frac{1}{2}J_{ll} + \frac{1}{2} J_{rr} - K_{lr} \\ 
W_{\text{HFB}}^{x = 0.5} = & J_{lr} - K_{lr}. 
\end{align} 
The difference in the Coulomb terms dictates the difference in behavior of the functionals. In case of $\lambda = 1$, all functionals essentially have non-zero Coulomb integrals whose self repulsion is cancelled 
by the exchange integral. This equivalence breaks down when $\lambda \neq 1$. The best example is $\lambda = 0$, for this value the diagonal Coulomb integrals will still generate a non-zero local 
self-repulsion, while the off-diagonal Coulomb integral only describes the long-distance repulsion. The exchange integral goes to zero, when the orbitals are fully localized. So, the LS functional at $\lambda = 0$ 
describes a system in which there is full local self-repulsion, and no long-distance repulsion, while the HFB and BB functionals only generate the physically correct long-distance repulsion and no local repulsion.

\subsection{Correcting homolytic dissociation for approximate functionals} 

We have seen that the approximate HFB and BB functionals have deficiencies when describing both the heterolytic and homolytic dissociation. The problems with the heterolytic dissociation are quite difficult to 
remedy, since one essentially has to completely cancel the symmetrical HFB/BB correction terms. For this scenario, one is essentially better off just using the HF functional from the start, if one knows that the 
system has an ionic description. If the system can typically undergo both heterolytic and homolytic dissociation depending on the circumstances it is recommended to refrain from using these approximate 
functionals. 

\begin{figure}[!p] 
\includegraphics[width= 0.5\textwidth, trim= 20 20 20 20]{\figdir/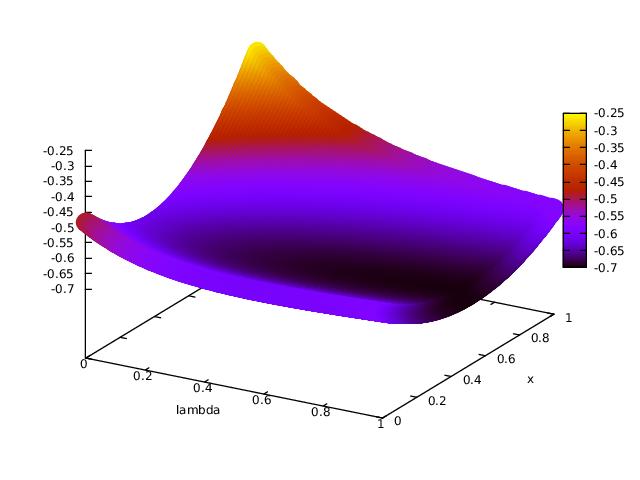} 
\caption{\label{fig:eHFBkappa} 
Hartree-Fock-Bogoliubov energy grid for $\frac{1}{R}$ = 0.2 hartree, using the one-electron $\lambda$-directing Lagrangian. 
The $\kappa$ value has been set to 1.} 
\end{figure}

The homolytic dissociation issues are, however, more easily fixable while still keeping the symmetrical correction terms, since one can force an energetically correct solution (on the $x = 0.5$ line) by applying 
restrictions during the optimization process in the spirit of constrained density functional theory \cite{KadukKowalczykVoorhis2012}. One option is to enforce $x = 0.5$ by adding a Lagrange multiplier expression 
that triggers when the occupation are close to $x = 0.5$, forcing them towards $x = 0.5$ and away from the physically incorrect minimum. Such a feat has already been performed in the past under the guise of 
equalizing the chemical potential on each atom \cite{ScuseriaTsuchimochi2009}. The downside of the approach is that it does not have a direct $\lambda$ directing effect, and the HFB and BB functionals 
themselves generate the same energy along the entire $x = 0.5$ line. If one wants to obtain natural orbital shapes that are equivalent to the exact ones ($\lambda = 1$), a $\lambda$-directing effect is 
mandatory. Such a solution would automatically force $x = 0.5$ as well, since it is the local minimum for $\lambda = 1$. The first choice that springs to mind is the LS two-electron energy expression 
$J_{ll} + J_{rr} - 2 K_{lr}$ at $x = 0.5$. However, apart from the fact that one might simply just use the entire LS functional, this expression is asymmetrical with respect integral prefactor signs. As a result, one 
cannot perform a quick integral transformation during the SCF process. Symmetrical two-electron option include the $\lambda$-less $J_{ll} + J_{rr} +2K_{lr}$ combination and the $J_{ll} + J_{rr} + 2J_{lr}$ 
combination, which have a non-zero contribution at $x = 0.5, \lambda = 1$. So, it seems impossible to find a suitable two-electron integral combination that is capable of forcing the system in the desired 
direction. The easiest way out is to use one-electron integrals instead. One can use 
\begin{align} 
\kappa (h_{ll} - h_{rr})^{2} = \kappa \left(\frac{(1 - \lambda)^{2}}{(1 + \lambda)^{2}}(h_{AA} - h_{BB})^{2}\right) 
\end{align} 
with $\kappa$ being a positive Lagrange multiplier. As is shown in the HFB plot for 5 bohr (Figure \ref{fig:eHFBkappa}), one can force the minimum in desired region using such an approach. In case of practical 
calculations, one has to take care that the correction only starts to play a role when the system starts to dissociate, since heteroatomic bonds do tend to have polarized charge distributions for equilibrium 
structures. So, one would most likely have to combine the orbital-directing and occupation-number-directing techniques for such scenarios. 

One could consider the addition of one electron components as an ad-hoc correction. However, one should see this in light of a broader potential direction which has not been pursued within DMFT. In DMFT the one electron part of the energy functional is known exactly, and an approximate functional has to be used for the two-electron component for systems containing more than 2 elecrons, since the general two-electron functional is unknown. However, the total outcome is the sum of the two components, and if one of them is (slightly) defunctional the final outcome will be less than optimal. One can either reject two-electron functionals yielding such an outcome, or add slight modifications of the one electron component in order to obtain a more correct outcome. The first option is of course preferred from a purist point of view, while the second option could be more interesting from computational efficiency point of view.

\section{Conclusion} 

In summary, we have analyzed the medium- to long-distance dissociation behavior of the HF, HFB, BB and LS functionals using a simple LiH model system. The results show that the HF functional tends to 
favor aufbau solutions which have more charge on the more electronegative atom (e.g., H atom in our case). In case of homolytic dissociation, such a solution is always incorrect since no charge imbalance 
should exist. The HF functional does generate the correct solution for the heterolytic scenario. The HFB functional adds a symmetrical (with respect to half occupancy of the orbitals) correction term to the HF 
functional. This correction term generates the correct homolytic dissociation energy at half occupancy ($x = 0.5$) of each orbital. However, the actual minimum is a compromise of this symmetrical factor and the 
HF solution, resulting in a minimum that still contains some small partial charge transfer. The same compromise also inhibits the correct description of the heterolytic dissociation in case the heterolytic global 
minimum is still energetically close to the homolytic minimum. The BB functional adds an additional term to the HFB functional, but this term is so small that there is essentially no meaningful difference between 
the HFB and BB functionals for our model. The LS functional is capable of handling both homolytic and heterolytic dissociation, which comes as no surprise since it is the exact functional for (singlet used here) 
two-electron systems. In case of homolytic dissociation, it forces the bonding and antibonding orbitals to be homopolar (same charge on each atom). For the (near) heterolytic scenario, it is able to create a 
corridor between the homolytic minimum and the global heterolytic ionic minimum due to the asymmetry that is present in its integral prefactors. The idea of forcing homopolar orbitals is used in a scheme in 
which a Lagrange multiplier expression forces the approximate HFB functional to such a minimum, resulting in correct energetic behavior and orbitals which are equivalent to the exact ones. 

Even though all of the calculations have been performed with DMFT functionals, many of the findings are most likely transferable to other cases in which one uses an approximate one-particle method in conjunction with a 
symmetrical particle-hole correction factor. A possible candidate is TAO-DFT. While TAO-DFT is formally exact for the ground-state electronic energy and density, approximate XC density functionals are typically 
adopted for practical TAO-DFT calculations. Therefore, it would be interesting to investigate how TAO-DFT with an approximate XC density functional performs for heteroatomic systems, and how the present 
Lagrangian correction scheme performs for reducing the errors (if any). We plan to pursue some of these issues in the near future.

\begin{acknowledgments} 
This work was supported by the Ministry of Science and Technology of Taiwan (Grant No.\ MOST107-2628-M-002-005-MY3), National Taiwan University (Grant No.\ NTU-CDP-105R7818), and the 
National Center for Theoretical Sciences of Taiwan. 
\end{acknowledgments}

\bibliography{rotationBIB.bib}

\newpage

\begin{table}[!p]
\centering
\begin{tabular}{l l | c c c} 
Functional	 & $\frac{1}{R}$ & $\lambda$ & $x$ & $E$ \\ 
\hline 
HF	&0	&0.26	&0.286	&-0.4949\\
	&0	&0.60	&0	&-0.4949\\
	&0.1	&0.51	&0	&-0.5490\\
	&0.2	&0.36	&0	&-0.6067\\
	&0.32	&0.01	&0	&-0.6945\\
	&	&	&	&\\
HFB	&0	&0	&0.438	&-0.7029\\
	&0.1	&0	&0.418	&-0.7050\\
	&0.2	&0	&0.382	&-0.7088\\
	&0.32	&0	&0.254	&-0.7227\\
	&	&	&	&\\
BB	&0	&0	&0.438	&-0.7029\\
	&0.1	&0	&0.418	&-0.7050\\
	&0.2	&0	&0.382	&-0.7088\\
	&0.32	&0	&0.254	&-0.7227\\
	&	&	&	&\\
LS	&0	&1	&0.500	&-0.6961\\
	&0.1	&1	&0.500	&-0.6961\\
	&0.2	&1	&0.500	&-0.6961\\
	&0.32	&1	&0.500	&-0.6961\\
	&0.32	&0	&0	&-0.6945\\
\end{tabular}
\caption{\label{tab:ModelMinima} 
The locations and values of the global (and a few local) minima for various $\frac{1}{R}$ (in hartree) and density matrix functionals.} 
\end{table} 

\end{document}